# MEIDNet: Multimodal generative AI framework for inverse materials design

Anand Babu[1,*], Rogério Almeida Gouvêa[1], Pierre Vandergheynst[2], Gian-Marco Rignanese[1,3,*]

[1]Institute of Condensed Matter and Nanosciences, Université Catholique de Louvain, Louvain-la-Neuve, Belgium.

[2]Signal Processing Laboratory 2, Institute of Electrical and Micro Engineering, School of Engineering, EPFL, Lausanne, Switzerland.

[3]WEL Research Institute, Avenue Pasteur 6, Wavre, Belgium

*Corresponding authors: anand.babu@uclouvain.be; gian-marco.rignanese@uclouvain.be

In this work, we present Multimodal Equivariant Inverse Design Network (MEIDNet), a framework that jointly learns structural information and materials properties through contrastive learning, while encoding structures via an equivariant graph neural network (EGNN). By combining generative inverse design with multimodal learning, our approach accelerates the exploration of chemical-structural space and facilitates the discovery of materials that satisfy predefined property targets. MEIDNet exhibits strong latent-space alignment with cosine similarity ≈ 0.96 by fusion of three modalities through cross-modal learning. Through implementation of curriculum learning strategies, MEIDNet achieves ~60 times higher learning efficiency than conventional training techniques. The potential of our multimodal approach is demonstrated by generating low-bandgap perovskite structures at a stable, unique, and novel (SUN) rate of 13.6 %, which are further validated by ab initio methods. Our inverse design framework demonstrates both scalability and adaptability, paving the way for the universal learning of chemical space across diverse modalities.

## 1. Introduction

The search for materials with desired properties is crucial for various applications, including energy storage, electronics, optoelectronics, and biomedical devices. However, conventional trial-and-error approaches are resource-intensive and have a limited scope. AI-enabled computational inverse design provides an efficient way to find candidates that satisfy predefined functional targets [1-3]. It exploits learned structure-property relationships to efficiently navigate complex chemical and structural landscapes. This approach significantly accelerates discovery cycles, guiding experimental efforts toward more targeted explorations [4-6].

Generative AI models, including variational autoencoders (VAEs) [7], generative adversarial networks (GANs) [8], and diffusion models [9-10] have demonstrated promising performance in the design and discovery of new materials. However, most of the proposed frameworks rely on a single mode of information, which limits their effectiveness in fully capturing the complex interplay among multiple property dimensions [10-13]. To address these limitations, multimodal machine learning (ML) has gained traction. By incorporating diverse sources of information, such as structural, electronic, mechanical, and thermodynamic properties, it facilitates the creation of a robust chemical latent space through shared learning [14-18].

Recent efforts have broadened the reach of multimodal frameworks by incorporating techniques such as contrastive learning [19], cross-modal attention mechanisms [20], and constrained-driven materials exploration [21]. For instance, the multimodal foundation model (MultiMat) integrates crystal structures, density of states (DOS), charge densities, and textual descriptions to uncover materials through latent-space fusion [22]. The composition-structure bimodal network (COSNet) improves the prediction accuracy for experimentally measured composition and structural data [23]. SCIGEN integrates geometric lattice constraints into a diffusion-based crystal generator to discover stable motif-guided quantum materials and validates a large subset via prescreening and DFT [21]. Similarly, denoising diffusion techniques coupled with cross-modal contrastive learning have enabled the guided discovery of chemical compositions and crystal structures from textual prompts [24]. Despite these initiatives, reliably generating stable, unique, and novel (SUN) materials with targeted properties and precisely navigating the latent space remains difficult. It requires extensive training and tighter alignment across modalities [15-16, 25]. Multimodality is clearly the future of materials science, yet only a handful of studies have addressed this challenge.

In this work, we presented a multimodal generative AI framework for inverse materials design named MEIDNet, which relies on contrastive learning among three modalities: structural, electronic, and thermodynamic properties. Structural encoding with a state-of-the-art equivariant graph neural network (EGNNs) is tested on three diverse datasets: Perovskite-5, MP-20, and Carbon 24. Latent-space alignment is evaluated for both early and late fusion strategies using the Information Noise-Contrastive Estimation (InfoNCE) loss function. We use the Perovskite-5 dataset (~19k five-atom $ABX_3$ structures) as a benchmark for our multimodal model. It is simple, widely used, and includes technologically relevant classes of materials, such as photovoltaics, ferroelectrics, and high-κ dielectrics. The cosine similarity and the L2 distances between modality embeddings are ~0.96 and ~0.24, respectively, indicating strong alignment. As a demonstration, 140 perovskite crystal structures are generated targeting thermodynamically stable materials in the low bandgap range (from 0.8 to 1.5 eV). The predicted bandgaps for the generated crystals closely match those determined with the crystal graph convolutional neural network (CGCNN) for single-property prediction (with an MAE of ~0.02). However, subsequent ab initio calculations show a lowering of the prediction. Out of the 140 generated perovskites, 19 are found to be stable, unique and novel, leading to a calculated SUN rate of ~13.6% without any further filtering, which is state of the art for multimodal models for materials to the best of our knowledge. Thus, our multimodal inverse design framework advances multimodal material generation efficiency and shows promise for scalability and adaptability by employing EGNNs for structural encoding and evaluating different fusion strategies. It leads the way toward universal learning of the chemical-structural space, accelerating the discovery of materials with desired properties, and propelling multimodal inverse design in materials science.

## 2. Results and Discussion

### E(3)-equivariant graph neural network and Multimodal framework

The crystal encoder transforms 3D crystal structures into an embedded latent representation using an EGNN for the structural encodings [26-27]. It is implemented by message passing [Eq. (i)], feature aggregation/update [Eq. (ii)], and coordinate update [Eq. (iii)]. It is equivariant to translations, rotations/reflections, and node permutations [28].

$$m_{ij} = \varphi_e\left(h_i^l, h_j^l, \left\|x_i^l - x_j^l\right\|^2, a_{ij}\right) \tag{i}$$

$$m_i = \Sigma_{j\neq i} m_{ij};\, h_i^{(l+1)} = \varphi_h\left(h_i^l, m_i\right) \tag{ii}$$

$$x_i^{(l+1)} = x_i^l + C \cdot \Sigma_{j \neq i}\left(x_i^l - x_j^l\right) \cdot \varphi_x\left(m_{ij}\right) \quad \text{(iii)}$$

$$z_s = \frac{1}{N}\sum_{i=1}^{N} h_i^{(L)}\left(m_{ij}\right) \quad \text{(iv)}$$

Here $x_i^l \in \mathbb{R}^n$ are node coordinates, $h_i^l \in \mathbb{R}^d$ node features, $a_{ij}$ edge attributes; $\mathrm{m_{ij}}$ the message (edge embedding) sent from node $j$ to node $i$, $\varphi_e, \varphi_x, \varphi_h$ multilayer perceptrons (MLPs), and $C$ a normalization constant. Finally, Eq. (iv) represents the global pooling operation where $h_i^{(L)}$ are the node features after the final EGNN layer L, N is the number of atoms in the unit cell, and $z_s$ is the resulting structural latent embedding.

The reconstruction fidelity of the autoencoder increases with the latent-space dimensionality, but so does the computational cost. As a result, a balance needs to be found. To assess this, we consider three latent dimensions, namely 64, 128, and 256, using the same number of training epochs on the Perovskite-5 dataset [29] as a benchmark. The 128-dimensional latent space emerges as the sweet spot, yielding a higher structure-matching (SM) rate (see Section 4. Methodology) at roughly the same computational cost (see Figure S1 in the Supplementary Information). Therefore, from here on, we use a 128-dimensional equivariant crystal autoencoder for multimodal alignment.

To validate the generalizability of our EGNN encoder, we also determine the reconstruction fidelity on two other diverse datasets: MP-20 [30] and Carbon-24 [31]. Thanks to the EGNN integrated autoencoder architecture, MEIDNet outperforms the unimodal Fourier transformed crystal properties (FTCP) [32] and crystal diffusion variational autoencoder (CDVAE) [7] for all three datasets in the SM rate (Table 1).

**Table 1**. Comparison of the reconstruction performance of MEIDNet, FTCP [32], and CDVAE [7] for the Perovskite-5 [29], MP-20 [30], and Carbon-24 [31] datasets.

| | Structure-matching rate (%) | | |
|---|---|---|---|
| Method | Perovskite-5 | Carbon-24 | MP-20 |
| FTCP | 99.34 | 62.28 | 69.89 |
| CDVAE | 97.52 | 55.22 | 45.43 |
| MEIDNet | 99.85 | 66.4 | 72.35 |

To address the latent search bottleneck in multimodality, we learn an aligned latent representation in which structural (crystal structure), electronic (bandgap), and thermodynamic (formation enthalpy) embeddings co-exist. As summarized in Figure 1a, the structural information is encoded through our EGNN. In parallel, we define a property encoder to map the scalar material properties (bandgap and formation enthalpy) into the shared latent space. This encoder is implemented as a Multilayer Perceptron (MLP) that projects the input property onto a 128-dimensional embedding. This ensures that both structural and property representations possess the same dimensionality, facilitating their alignment via contrastive learning.

For robust joint learning, we investigate the effect of early and late fusion approaches on multimodal alignment (Figure 1b, more details are given in Section S1 of the Supplementary Information). In early fusion, modality-specific features are merged at the input feature level, and a shared network learns a joint representation. In late fusion, each modality is encoded separately, and the resulting embeddings are combined at the alignment stage. Contrastive learning unifies structural and property encodings in a joint latent space, which facilitates interactions between distinct modalities and optimizes the alignment of their information [19-20].

The alignment between modalities is achieved via contrastive training using the InfoNCE loss [33-34]

$$\mathcal{L}_{\text{InfoNCE}} = -\frac{1}{B}\sum_{k=1}^{B}\log\frac{\exp\left(\text{sim}\left(\boldsymbol{z}_{\boldsymbol{s}}^{(\boldsymbol{k})},\boldsymbol{z}_{\boldsymbol{p}}^{(\boldsymbol{k})}\right)/\tau\right)}{\sum_{l=1}^{B}\exp\left(\text{sim}\left(\boldsymbol{z}_{\boldsymbol{s}}^{(\boldsymbol{k})},\boldsymbol{z}_{\boldsymbol{p}}^{(\boldsymbol{l})}\right)/\tau\right)} \quad \text{(v)}$$

where $B$ is the batch size, and indices $(k)$ and $(l)$ denote samples within the mini batch. $z_s^{(k)}$ and $z_p^{(k)}$ are the aligned structural and property embeddings for the $k$-th crystal, $sim(u, v) = u^{\mathrm{T}}v/(\|u\|\|v\|)$ denoting cosine similarity. τ is the temperature hyperparameter, and the denominator sums over all $l$ samples in the batch to normalize the probability.

We implement an inverse design pipeline for target-led navigation in the aligned latent space. An iterative optimization is performed until the predicted properties converge to the targeted values for the generated crystal structure (Figure 1c). Thus, MEIDNet offers a more compact latent space than diffusion models, facilitating interpretability and navigation. In addition to this, it is scalable to numerous modalities toward a unified latent representation.

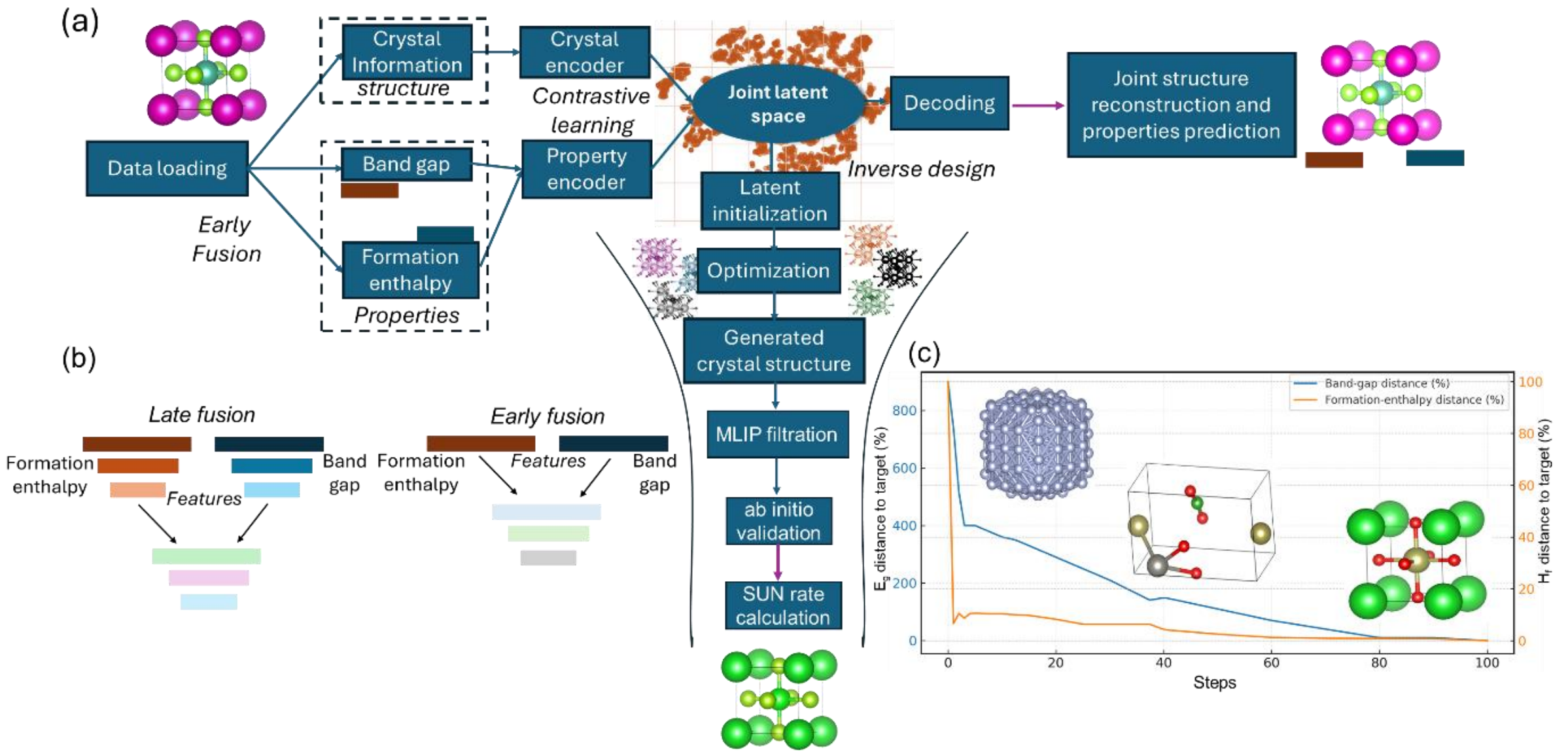

**Figure 1. Workflow schematics of the MEIDNet framework for inverse materials design**: (a) The process starts with data loading, extracting crystal information and corresponding properties (here, the band gap and formation enthalpy) which are fused. These two modalities are encoded and aligned into a joint latent space through contrastive learning, where iterative optimization takes place with respect to the target properties. The optimized latent representations are then decoded into crystal structures, which are then filtrated using MLIPs and validated ab initio. (b) Different fusion approaches are considered: early fusion merges modality-specific features at input so a shared network learns a joint representation, whereas late fusion encodes modalities separately and combines their embeddings during alignment. (c) Generation of structures through navigation on the aligned latent space, where the *y*-axis represents the distance from the target properties.

## Curriculum learning-enabled training and multimodal fusion strategies

In order to improve the learning efficiency, we employ a progressive weighting on the contrastive loss: its weight $\alpha_t$ is ramped up from 0 to 1 over the first two thirds of training epochs and then held at 1, improving reconstruction, latent alignment, and convergence [35-36]. The pacing follows an exponential schedule:

$$\alpha_t = 1 - \exp(-\gamma t) \quad \text{(vi)}$$

where $t$ is the epoch index and $\gamma$ controls the ramp rate. This training strategy leads to 60-times faster learning with respect to traditional training (Figures 2a and 2b).

We thoroughly investigate the effect of different fusion strategies on the latent space alignment: late fusion (LF), early fusion (EF), and early fusion with curriculum learning (EF+CL). We adopt three key metrics: cosine similarity (CS), structure-matching (SM) rate, and L2 distance (L2D) on the scale of 0-1, 0-100 % and 0-2, respectively. To determine the best strategy, we train MEIDNet on the Perovskite-5 dataset using 200 epochs. The results are illustrated in Figure 2a. LF demonstrates a good SM performance (~66%) but performs badly in latent space alignment (CS~0.31, L2D~1.06). In contrast, EF displays a good alignment (CS~0.89, L2D~0.45), but a lower SM performance (~32%). This can likely be explained by the premature mixing of heterogeneous descriptors causing feature interference/over-smoothing, which yields a measurable drop versus LF, especially in reconstruction fidelity. EF+CL outperforms the previous two fusion strategies: it has a SM performance (~66%) similar to LF and a better alignment (CS~0.91, L2D~0.4) than EF. This promising performance is due to the initial stage of training CL giving more weight to learning the crystal structures, while their corresponding properties are scalars, which are easier to learn at a later stage [37].

For production, MEIDNet is then trained on the same Perovskite-5 dataset but using 2000 epochs and adopting the EF+CL strategy. It demonstrates a high cosine similarity (CS~0.97), which suggests that the shared latent space effectively captures the inherent relationships between structural motifs and their corresponding properties [38-39]. In addition to the cosine similarity, the latent representations exhibit an average L2 distance (L2D) of ~0.24, which indicates that the multimodal embeddings are closely aligned in the latent space [40]. To assess joint learning, we perform regression analyses showing that the jointly reconstructed SM score (Figure 2c), bandgap ($E_g$) (Figure 2d), and formation enthalpy ($H_f$) (Figure 2e) track the ground truth with near-linear trends ($R^2 \approx 0.996$). This consistency indicates that the shared representation captures cross-modal correlations and reduces prediction variance.

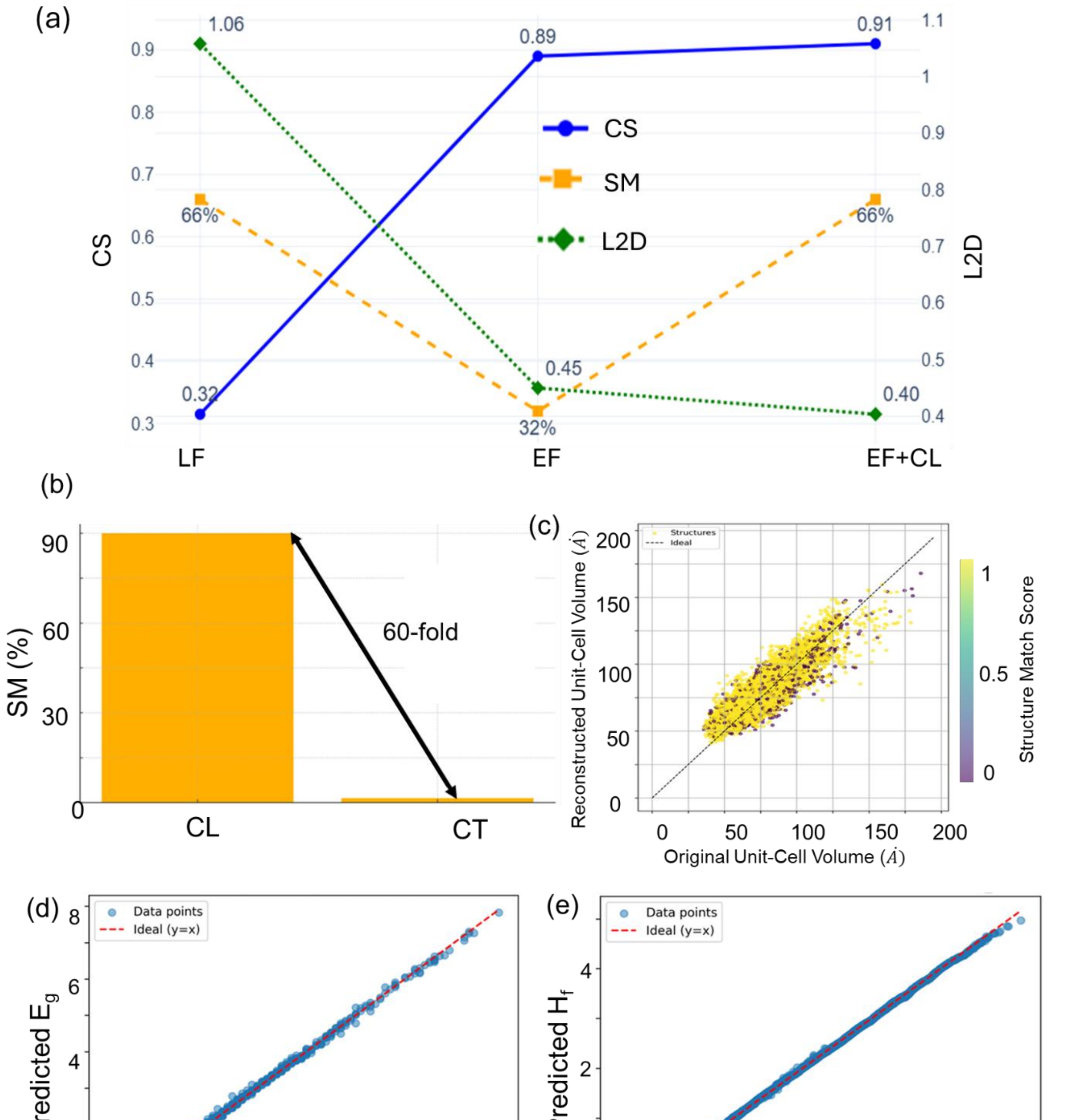

**Figure 2. Evaluation of multimodal machine learning model efficiency and joint reconstruction accuracy**: (a) Comparative evaluation of multimodal machine learning performance metrics: Performance analysis of late fusion (LF), early fusion (EF) and EF with curriculum learning (EF+CL). The performance metrics are through cosine similarity (CS), structure matching (SM), L2 distance (L2D) and $R^2$ metrics, which highlights the advantages of EF+CL in enhancing latent alignment and reducing prediction errors (b) 60-fold improvement in training efficiency with curriculum learning (CL) versus conventional training (CT), as evidenced by higher SM (%) at a fixed number of epochs. (c) Scatter plot demonstrating structure reconstruction and structure matching score prediction of (d) bandgap ($E_g$) and (e) formation enthalpy ($H_f$) for MEIDNet training.

**Inverse design-led targeted structure generation**

To demonstrate the inverse-design pipeline, we test MEIDNet for generating perovskites structures (details provided in Section 4. Methodology). To evaluate target-led generation, we perform conditional sampling in the low bandgap range ($E_g \in \{0.8, 1.0, 1.2, 1.5\}$ eV) under a fixed negative formation enthalpy of $-25$ meV. For each target, a batch of 35 candidate materials are produced by the model.

The thermodynamic stability of the generated candidates is first assessed using the eSEN-30M-OAM machine learning interatomic potentials developed by Facebook AI Research (FAIR) [41] at Meta (Figure 3a). One of the key metrics to quantify the success of generative models in materials science is the fraction of stable, unique, and novel structures, the so-called SUN rate. We adopt the same conventional thermodynamic-stability criterion used in previous generative models, which consider that structures with energy above hull ($E_{hull}$) below 100 meV, as calculated from the Materials Project database, are stable [7,9,32,43]. Out of 140 generated structures, 19 meet this criterion and are not present in the training dataset, yielding a SUN rate of approximately 13.6% (Figure 3b, Table S1, and Figure S2, Supplementary Information). These findings are further confirmed by ab initio calculations using the Vienna Ab initio Simulation Package (VASP) (see Methodology). We highlight that, although the model is trained mainly on unstable high $E_f$ perovskites (Figure 2e), it still reliably leads to thermodynamically stable structures.

This places our model at the forefront of inverse-design-based multimodal learning approaches with respect to its counterparts [32, 42]. To the best of our knowledge, our model trained on multiple modalities is state-of-the-art in terms of the SUN rate (without using any filtration) for multimodal models for materials. It is even competitive with generative models trained on single modality including MatterGen ~39% [9], CDVAE ≈ 18% [7], FTCP < 5% [32], and Matra-Genoa~16 % [43].

Turning to the validation of the target bandgap, we first use the CGCNN predictor and obtain an MAE of ~0.02 eV as detailed in the Supplementary Information (Figure S3 and Table S1). This shows that the predictive capabilities of MEIDNet are comparable to those of the leading GNN models for predictions. The bandgaps are also computed at the PBE level (without +U or SOC). The calculated values are often lower than the requested target. This downward shift likely reflects (i) a modest training-set skew toward smaller gaps in the Perovskite-5 dataset and (ii) the expected low coverage in the training set for the novel perovskite compositions (iii) band structures were computed at the PBE level without +U.

In terms of the generated structures, we first would like to highlight four novel structures with a bandgap in the low range and very interesting properties (see Figure 3c): (i) $YbScSe_3$, (ii) $LaScSe_3$, (iii) $BaHfSe_3$ and (iv) $KTaSe_3$. Their corresponding $E_{hull}$ *values* are 0, 91.88, 66.78, and 88.6 meV, respectively, when calculated with eSEN and 0, 95.6, 98.7, and 0 meV, respectively, when calculated within DFT. Other promising perovskites found using our workflow are presented in the Supplementary Information (Figures S5 and S6, and Table S1).

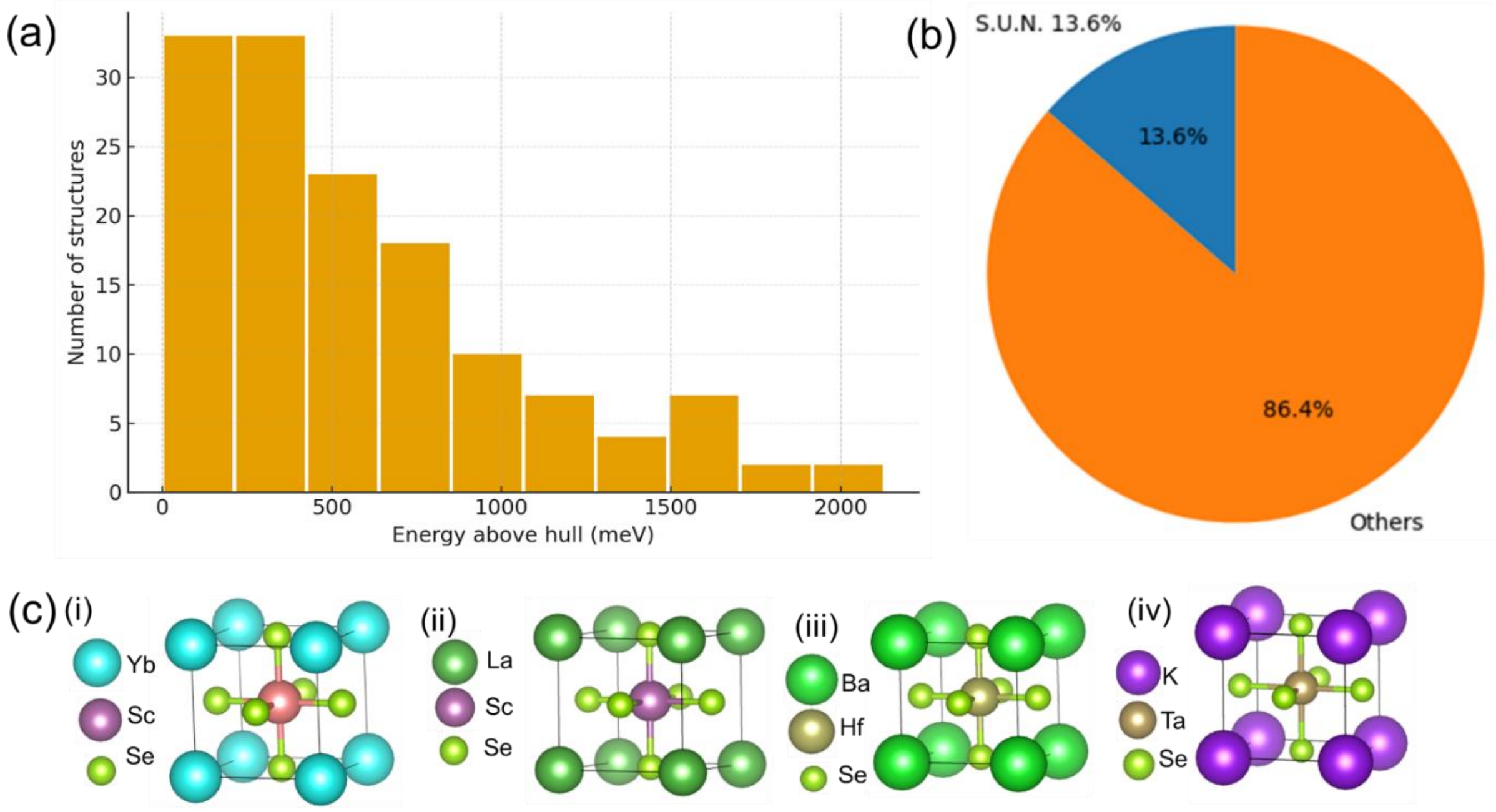

**Figure 3. Analysis and validation of optimized structures generated by MEIDNet:** (a) Distribution of energies above the convex hull for generated structures. (b) Stable, unique and novel (SUN) rate indicating the potential of target led materials discovery with MEIDNet. (c) Crystal structure of SUN perovskite materials generated by MEIDNet at the low bandgap range and formation enthalpy -25 meV/atom (c) (i) $YbScSe_3$ (ii) $LaScSe_3$ (iii) $BaHfSe_3$ (iv) $KTaSe_3$.

$YbScSe_3$ shows broadly dispersive bands with a shallow indirect separation between the valance and conduction bands and a relatively smooth density of states (DOS) around the gap (Figure 4a). This suggests good baseline mobility and temperature-activated transport. Additionally, strain or epitaxial constraints may offer a potential pathway to tune gap directness or drive band inversion in these materials. $LaScSe_3$ (Figure 4b) exhibits a narrow, weakly dispersive conduction manifold just above the Fermi level, producing a sharp DOS onset and implying heavy electron effective masses in certain directions, whereas the valence edge is more dispersive and consistent with lighter holes. The indirect gap with a flat plus dispersive topology is attractive for further investigations in its thermoelectric properties and for strain tunable band edge engineering. The band structure of $BaHfSe_3$ (Figure 4c) indicates an indirect very small-gap semiconductor. The valence band structure near R implies light holes, equally the dispersive CBM at Γ signals light electrons. $KTaSe_3$ (Figure 4d) shows a metallic band structure, with finite DOS at Fermi energy and steep dispersions, pointing to intrinsically good conductivity. Thus, the results indicate that MEIDNet can generate physically plausible structures within the requested targets of low bandgap and thermodynamic stability.

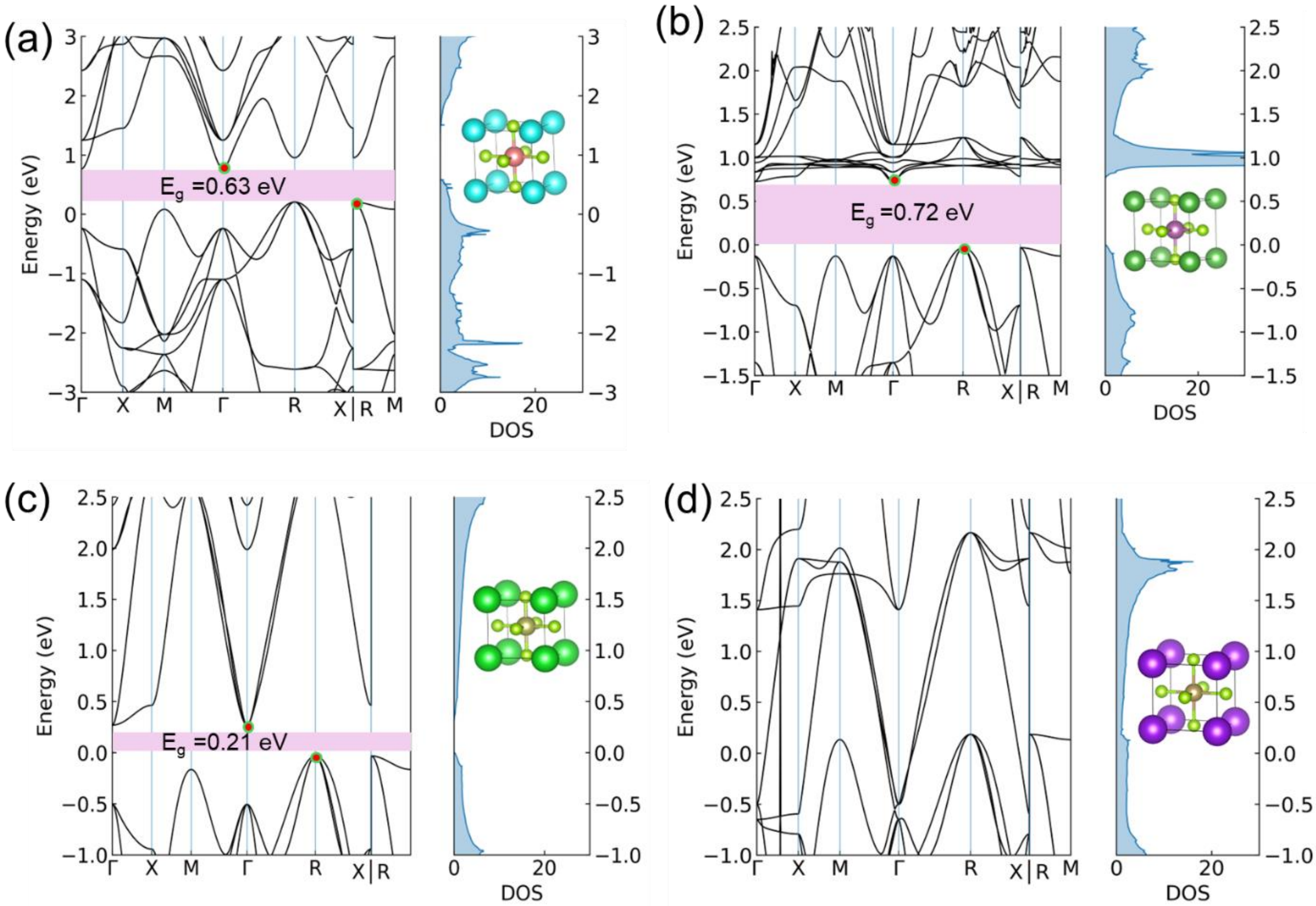

**Figure 4. Band structure and density of states (DOS) plots of selected SUN materials generated by MEIDNet**: (a) $YbScSe_3$ (b) $LaScSe_3$ (c) $BaHfSe_3$ (d) $KTaSe_3$ at the low bandgap range target with the constant formation enthalpy of -25 meV/atom.

Our framework addresses key prerequisites by first identifying thermodynamically plausible candidates that satisfy the conventional energy-above-hull criterion against known phases. However, we must also confirm that the material's crystal structure is dynamically stable, meaning that any small thermal vibration or perturbation would not be enough to cause it to spontaneously transform into a different structure.

Therefore, we probe the dynamical stability of the novel perovskites obtained from MEIDNet by computing phonon dispersions using Phonopy with the MACE-OMAT-0 model [44]. Our analysis reveals that these structures, despite their thermodynamic stability, exhibit soft modes in their phonon dispersions, indicating that they are dynamically unstable at 0 K. Although it is possible that these instabilities could be resolved at finite temperatures, we opt to provide a more immediate path to viable materials.

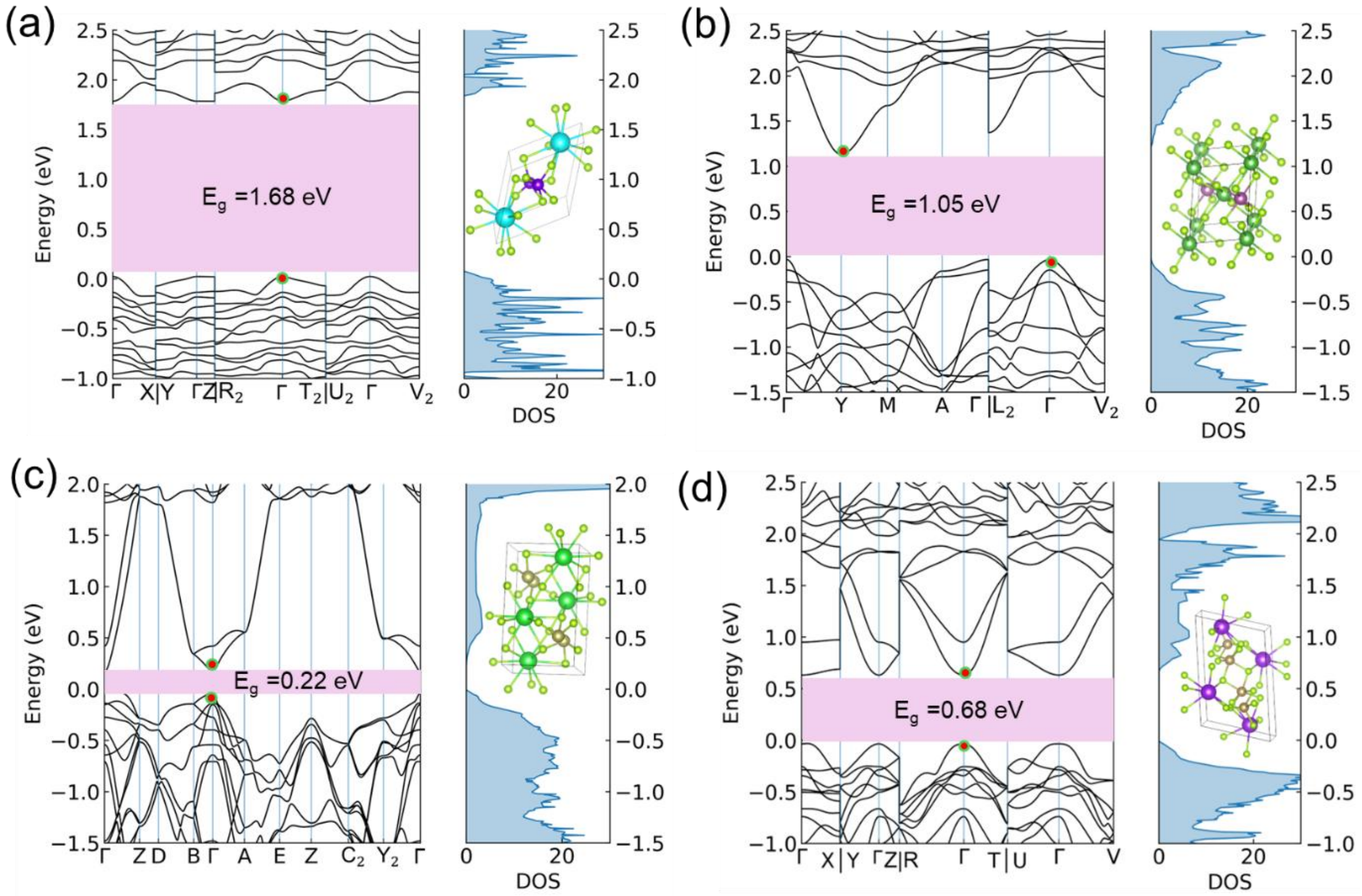

**Figure 5. Band structure of the VibroML-treated SUN materials at low bandgap energy**: (a) $YbScSe_3$ (b) $LaScSe_3$ (c) $BaHfSe_3$ (d) $KTaSe_3$.

We utilize the VibroML toolkit (see Section 4 Methodology) to follow these soft modes, systematically mapping the potential energy surface to identify the most stable, corresponding polymorphs. The verified convex hull distances for each of the lowest energy polymorphs of VibroML treated $YbScSe_3$, $LaScSe_3$, $BaHfSe_3$, $KTaSe_3$ are 0, 58, 0 and 0 meV, respectively, indicating the reduction in energy as shown in Supplementary Information, Table S2. The lower-energy polymorphs found show larger bandgaps than the original pristine perovskite structure, as shown in Figure 5. This is expected, as these polymorphs exhibit a reduced electronic dimensionality compared to the 3D-connected pristine phase. The conduction and valence bands of the low-energy polymorphs of $YbScSe_3$, $LaScSe_3$, and $BaHfSe_3$ are generally less dispersive than their pristine perovskite counterparts, but given their lower energy and small bandgaps, these materials remain promising for further investigation for diverse optoelectronic applications. Compared to the metallic pristine $KTaSe_3$ structure, the corresponding polymorph found presents a small direct bandgap of 0.68 eV with fairly dispersive conduction and valence bands at Γ, which makes this a very promising material for optoelectronic applications.

Furthermore, the band structures are calculated for other MEIDNet-generated structures that pass the thermodynamic and dynamic stability screening (see Table S2 and Figure S7). These are found to show similar promising results, as presented in Figures S5 and S6 in Supplementary Information. To facilitate further investigation and collaboration within the community, the CIFs of all the novel structures found with the MEIDNet and VibroML workflow are provided in our Support Data repository.

In summary, this work establishes that integrating multimodal generative models like MEIDNet with dynamic instability remediation and ab initio validation is a highly promising strategy. Our combined workflow (MEIDNet-VibroML-DFT) effectively delivers a curated set of physically grounded materials, providing the research community with a curated set of promising, physically grounded materials for experimental consideration that can be easily extended for other systems of interest.

## 3. Conclusions

We introduced MEIDNet, a multimodal inverse-design framework that jointly learns structural information and materials properties through latent space alignment. We here focused on electronic and thermodynamic properties, but the model can be easily adapted to other properties. Our autoencoder-based multimodal framework offers more interpretable, easily navigable learning space, compared with diffusion models, and it is naturally extensible to additional modalities toward a universal latent representation. By coupling an equivariant graph neural network with curriculum-guided contrastive training, the model achieves faster convergence and improved latent alignment.

We demonstrated the utility of our end-to-end framework by generating novel perovskite candidates with pre-determined properties with a competitive SUN rate. More broadly, this success demonstrates a powerful and generalizable framework poised to accelerate the computational discovery of stable materials with targeted properties across diverse chemical classes.

## 4. Methodology

### Dataset Preparation and Representation

The training and evaluation datasets consist of crystal structures originating from the Perovskite-5 (~19k), MP-20 (~45k), and Carbon-24 (~10k) datasets. The crystal structures and properties are converted into a dense numerical representation suitable for machine learning using the Python Materials Genomics (pymatgen) library. Lattice parameters are encoded by EGNN, while atomic species are represented using one-hot encoding across a comprehensive list of 56 elements (detailed information is provided in Section S2 of the Supplementary Information).

### Model Architecture and Training

We develop a dual and tri-auto encoder architecture integrating a crystal graph autoencoder with a property encoder-decoder network. The crystal autoencoder is employed through EGNN layers capable of encoding spatial geometry and atomic species information. The model utilizes three sequential EGNN layers, enabling accurate encoding and decoding of both spatial coordinates and adjacency matrices representing atomic interactions.

To train the model, a contrastive learning approach is adopted to align the crystal structure and scalar property modalities in a shared latent space. The training loss function combines reconstruction losses for the structural descriptors (lattice parameters, adjacency matrices, atomic species, and atomic coordinates) with mean squared error (MSE) for scalar property predictions. All training procedures employ the Adam optimizer with a learning rate of $1e^{-3}$, a batch size of 16, and training extends up to 2000 epochs. Training is carried out on GPU-equipped high-performance computing clusters to facilitate rapid convergence. All computational workflows are implemented using Python, with extensive use of deep learning frameworks (PyTorch), materials informatics libraries (pymatgen), and numeric analysis tools (numpy, pandas). Code execution is carried out via GPU-based high-performance computational environments.

**Inverse Design Framework, SUN screening and ab initio validation**

The inverse-design pipeline is implemented as follows. Randomly initialized vectors in the aligned latent space are gradually improved via gradient descent using the Adam optimizer. During this navigation in the latent space, the vectors move toward structures with the target properties (bandgap and formation enthalpy). Optimization is guided by weight losses incorporating physical constraints, such as lattice volume, minimum interatomic distances, and charge neutrality.

We also probe how adding domain constraints affects inverse design by comparing models trained with early constraints against models trained without them. These constraints, such as enforcing charge neutrality, ensuring valid oxidation states, maintaining minimum interatomic distances, and using tolerance factor windows, steer the model effectively minimizing the impact in the exploration of novel structures. As shown in Figure S4, in Supplementary Information, introducing constraints early leads to faster convergence and better reconstruction quality with fewer epochs, however, it narrows down the exploratory breadth [45]. Regardless of the training regime, applying constraints during generation helps guide the decoded structures toward physically plausible crystals [46].

The generated crystal structures undergo structural relaxation using the eSEN-30M-OAM machine learning interatomic potential (MLIP) to minimize internal forces and energies, achieving physically plausible equilibrium configurations. The relaxation convergence criteria include a force threshold of 0.01 eV/Å and a maximum optimization step count of 300. Using the Materials Project API, the energy above hull is evaluated to find the targeted stable generated structures. Stable structures are further validated by the Ab Initio calculations using VASP with PBE PAW pseudopotentials. For the latter calculations, we adopt the Materials Project default parameters: a 520 eV plane-wave cutoff and a k-point density of 1000 k-points per reciprocal atom (KPPRA). These standardized settings place total energies on a consistent scale, enabling reliable convex-hull construction and thermodynamic stability assessment [47-49].

**Dynamical stability evaluation**

The dynamical stability is investigated using Phonopy [50] and VibroML [51]. The latter is an automated toolkit for identifying and stabilizing dynamically unstable crystalline phases using machine-learned interatomic potentials (MLIPs) [52]. The algorithm employs a two-stage optimization approach. First, a parameter sweep optimization identifies optimal phonon calculation settings (supercell size, atomic displacement, force tolerance) that minimize imaginary phonon frequencies. When soft modes persist, the toolkit automatically triggers a genetic-algorithm-driven structure exploration that generates new atomic configurations by displacing atoms along unstable phonon eigenmodes with varying displacement scales, mode coupling ratios, and cell transformations.

The genetic algorithm (GA) evolves a population of structural variants through selection, crossover, and mutation operations, where fitness is determined by the relaxed energy per atom of each generated structure. Each GA generation produces offspring structures that are relaxed using MLIPs, and the algorithm iteratively refines the search space by performing phonon analysis on the most promising candidates to identify new soft modes for subsequent iterations. This approach enables systematic exploration of the potential energy landscape to discover lower-energy, dynamically stable phases from initially unstable starting structures.

For the generated perovskites, a total of 19 generations are evaluated, with all calculations performed using the MACE-OMAT-0 model [46-53]. From the three structures with the lowest energy and no soft modes generated by the workflow, we select the one with the highest symmetry for subsequent calculations.

## Data availability

The crystal structures generated via MEIDNet and implementation code are available at https://github.com/ABnano/MEIDNet.

## Acknowledgements

We gratefully acknowledge high-performance computing support from the Université Catholique de Louvain (CISM/UCL) and the Consortium des Équipements de Calcul Intensif en Fédération Wallonie-Bruxelles (CÉCI). This work also relied on the use of Lucia, the Tier-1 supercomputer of the Walloon Region.

## Funding

This study was financed by Université catholique de Louvain (Ref. No.: ARH/MKK/01155003).

### Author contributions

A.B., R.G., and G.M.R. conceptualized the study. A.B. performed the experiments and data analysis and wrote the main manuscript. R.G. carried out vibrational analysis. P.V., and G.M.R. supervised the investigation. A.B., R.G., P.V., and G.M.R. reviewed and approved the final manuscript.

### Competing interests

The authors declare no competing interests.

## Supplementary Information

# MEIDNet: Multimodal generative AI framework for inverse materials design

Anand Babu[1,*], Rogério Almeida Gouvêa[1], Pierre Vandergheynst[2], Gian-Marco Rignanese[1,3,*]

[1]Institute of Condensed Matter and Nanosciences, Université Catholique de Louvain, Louvain-la-Neuve, Belgium.

[2]Signal Processing Laboratory 2, Institute of Electrical and Micro Engineering, School of Engineering, EPFL, Lausanne, Switzerland.

[3]WEL Research Institute, Avenue Pasteur 6, Wavre, Belgium

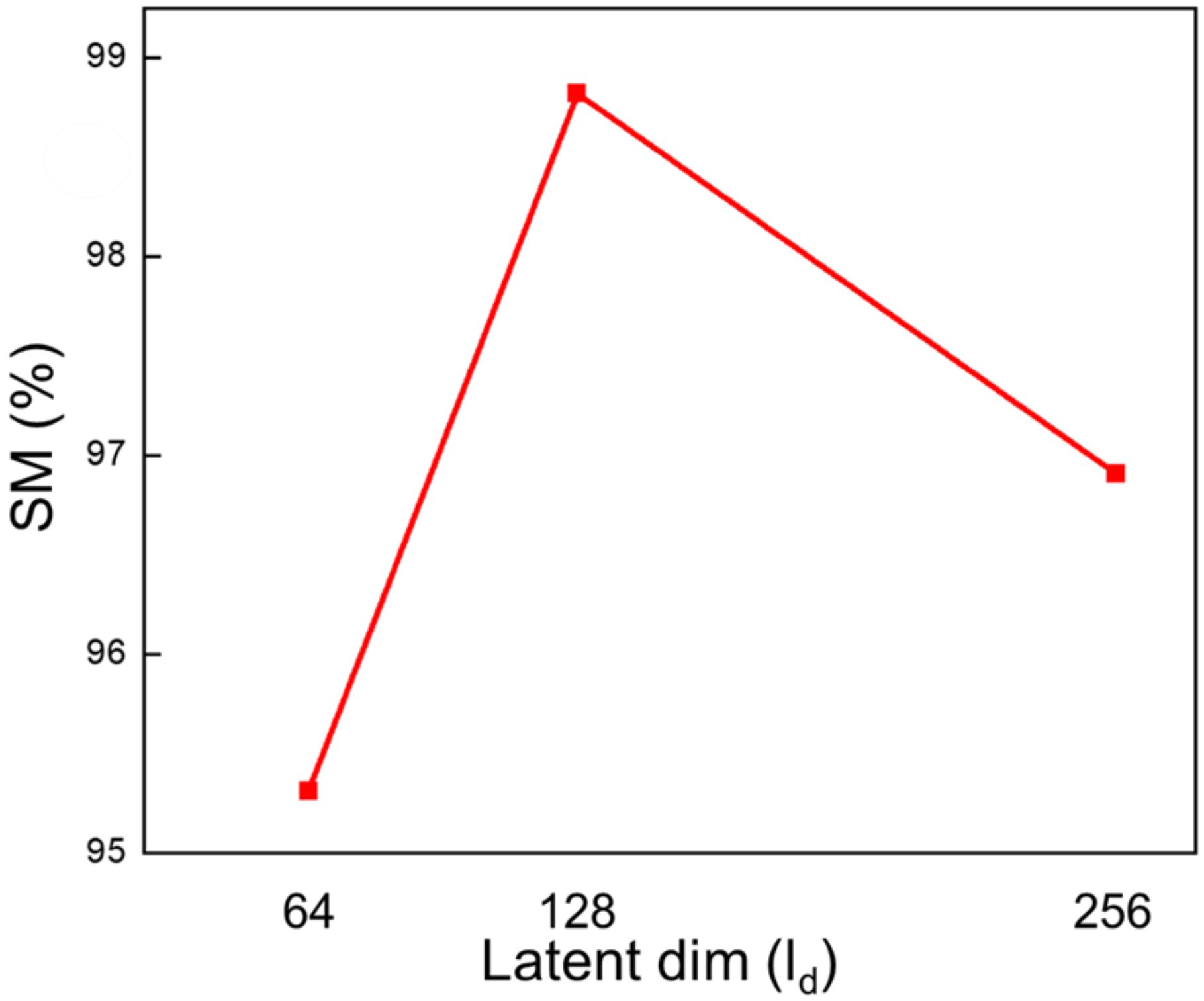

**Figure S1.** Latent space dimensionality sweeps of EGNN under a fixed 500 epochs training schedule, using reconstruction loss and structure matching (SM). A 128-dimensional latent provides the best SM performance, which is adopted for MEIDNet training.

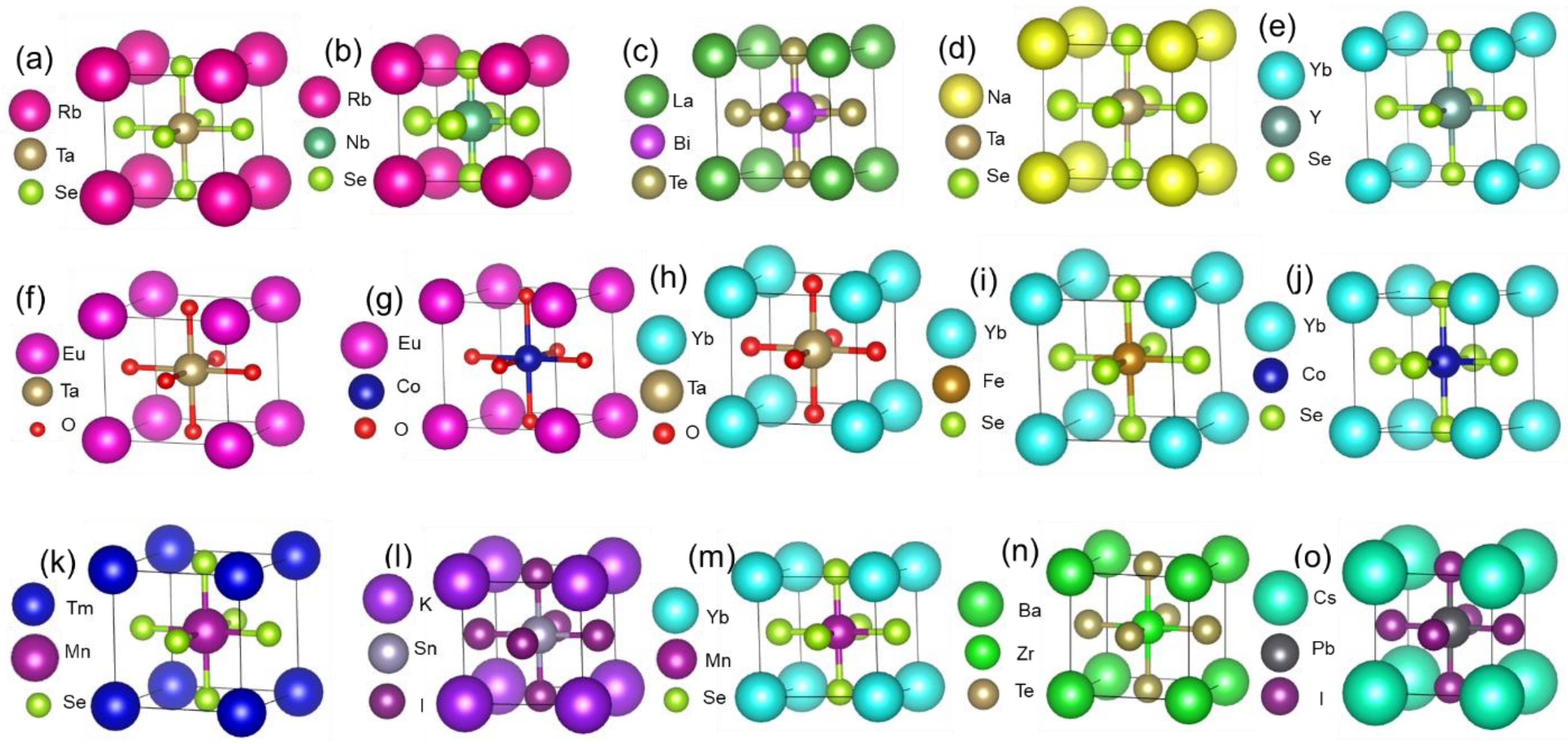

**Figure S2**. More SUN materials with diverse targets generated by MEIDNet at bandgaps of 0.8 eV (a) $RbTaSe_3$ (b) $RbNbSe_3$, (c) $LaBiTe_3$ (d) $NaTaSe_3$.1 eV: (e) $YbYSe_3$ (f) $EuTaO_3$. 1.2 eV: $EuCoO_3$ (g) $YbTaO_3$ (h) $YbFeSe_3$ (i) $YbCoSe_3$ (j) $TmMnSe_3$. 1.5 eV: (k) $KSnI_3$ (l) $YbMnSe_3$ (m) $BaZrTe_3$ (n) $CsPbI_3$ at constant formation enthalpy of -25 meV/atom.

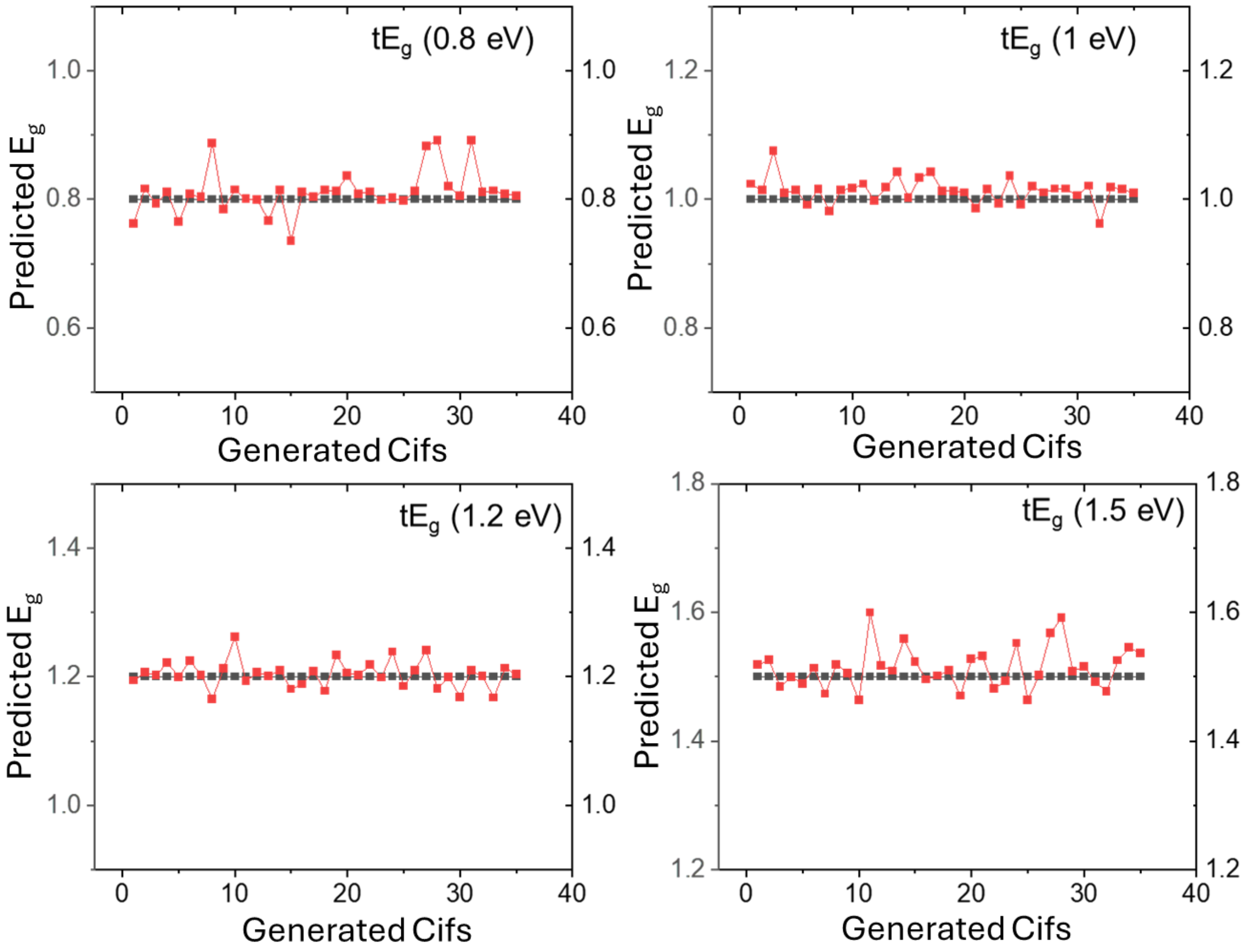

**Figure S3.** Evaluation of predictive accuracy of the MEIDNet with CGCNN. (a) with the target bandgap of ($tE_g$) (a) 0.8 eV, (b) 1 eV, (c) 1.2 eV and (d) 1.5 eV at constant formation enthalpy of -25 meV/atom.

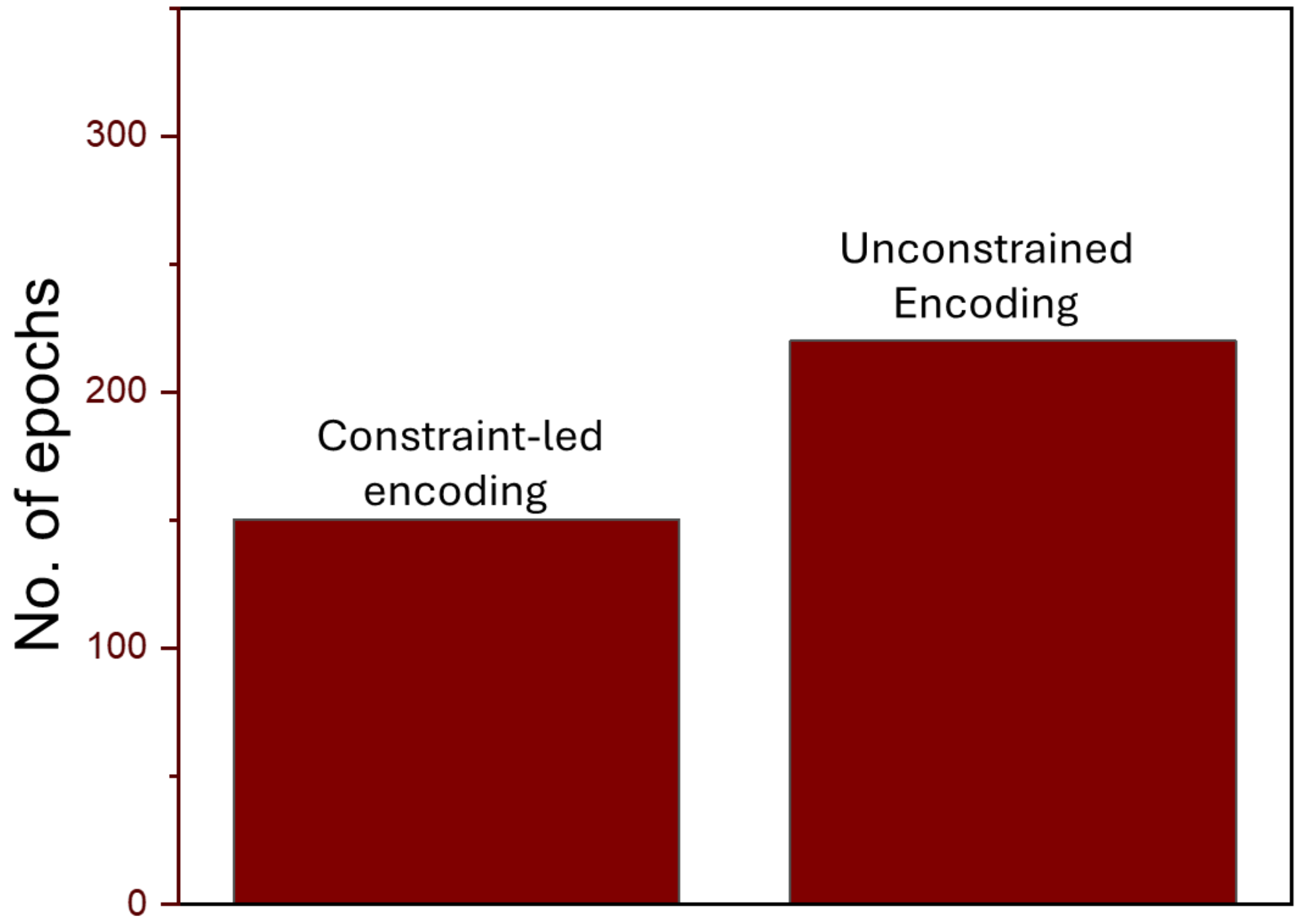

**Figure S4**. Number of training epochs required to attain a structure-matching accuracy of SM = 96.8% for MEIDNet trained under two regimes: constrained-led encoding (chemistry/geometry constraints applied) and unconstrained (unconstrained encoding). The domain constrained model reaches the target SM in fewer epochs than the unconstrained baseline, which indicate faster optimization under the same training schedule.

**Table S1.** The SUN materials generated by MEIDNet at constant formation enthalpy of -25 meV/atom; here, $E_g$ [Target] represents the target bandgap; $E_g$ [CGCNN] is bandgap predicted by CGCNN; $E_g$ [DFT] is bandgap calculated by DFT; $E_{hull}$ [eSEN] is energy above hull predicted by the eSEN model and $E_{hull}$ [DFT] is the energy above hull calculated by DFT.

| S. No. | Formula | $E_g$ [Target] (eV) | $E_g$ [CGCNN] (eV) | $E_g$ [DFT] (eV) | $E_{hull}$ [eSEN] (meV) | $E_{hull}$ [DFT] (meV) |
|---|---|---|---|---|---|---|
| 1 | $RbTaSe_3$ | 0.8 | 0.792 | 0 | 98.81 | 94.2 |
| 2 | $RbNbSe_3$ | 0.8 | 0.81 | 0 | 49.35 | 97.6 |
| 3 | $LaBiTe_3$ | 0.8 | 0.764 | 0.14 | 94.61 | 0 |
| 4 | $KTaSe_3$ | 0.8 | 0.807 | 0 | 88.6 | 0 |
| 5 | $NaTaSe_3$ | 0.8 | 0.784 | 0.91 | 92.3 | 0 |
| 6 | $LaScSe_3$ | 1 | 0.981 | 0.72 | 91.88 | 95.6 |
| 7 | $EuTaO_3$ | 1 | 1.016 | 0 | 0 | 0 |
| 8 | $YbYSe_3$ | 1 | 1.023 | 0.78 | 0 | 0 |
| 9 | $YbScSe_3$ | 1 | 0.997 | 0.63 | 0 | 0 |
| 10 | *$EuCoO_3$ | 1.2 | 1.192 | 0 | 28.4 | 0 |
| 11 | *$YbTaO_3$ | 1.2 | 1.206 | 0 | 97.6 | 0 |
| 12 | $YbFeSe_3$ | 1.2 | 1.201 | 0 | 0 | 0 |
| 13 | $YbCoSe_3$ | 1.2 | 1.211 | 0 | 0 | 0 |
| 14 | $TmMnSe_3$ | 1.2 | 1.18 | 0 | 81.8 | 88.6 |
| 15 | *$KSnI_3$ | 1.5 | 1.501 | 1.89 | 0 | 0 |
| 16 | $YbMnSe_3$ | 1.5 | 1.509 | 0 | 0 | 0 |
| 17 | $BaZrTe_3$ | 1.5 | 1.47 | 1.1 | 96.3 | 0 |
| 18 | *$CsPbI_3$ | 1.5 | 1.526 | 1.48 | 28.43 | 25 |
| 19 | $BaHfSe_3$ | 1.5 | 1.48 | 0.21 | 66.78 | 98.7 |

* Marks materials available in the Materials Project database but not in the training data.

**Table S2.** SUN materials treated by VibroML for dynamical stability. Here, $E_{hull}$ [DFT] is the energy above hull of the SUN materials calculated by DFT; $E_{hull}$ [VibroML+DFT] is the energy above hull calculated by DFT of the SUN materials treated by VibroML. $E_g$ (DFT) is the band gap calculated by the DFT; $E_g$ (VibroML+DFT) is the bandgap calculated by the DFT of the SUN materials treated by VibroML.

| S. No. | Formula | $E_{hull}$ [DFT] (meV) | $E_{hull}$ [VibroML+DFT] (meV) | $E_g$ [DFT] (eV) | $E_g$ [VibroML+DFT] (eV) |
|---|---|---|---|---|---|
| 1 | $KTaSe_3$ | 0 | 0 | 0 | 0.68 |
| 2 | $LaScSe_3$ | 95.6 | 58 | 0.72 | 1.05 |
| 3 | $YbScSe_3$ | 0 | 0 | 0.63 | 1.68 |
| 4 | $LaBiTe_3$ | 0 | 0 | 0.14 | 1.38 |
| 5 | $YbFeSe_3$ | 0 | 0 | 0 | 0.79* |
| 6 | $YbMnSe_3$ | 0 | 0 | 0 | 1.98* |
| 7 | $YbCoSe_3$ | 0 | 0 | 0 | 0.67* |
| 8 | $TmMnSe_3$ | 88.6 | 41 | 0 | 0.79* |
| 9 | $BaHfSe_3$ | 98.7 | 0 | 0.21 | 0.22 |

* Values marked with an asterisk correspond to structures predicted as metallic. The shown gaps are estimated assuming a 1-2 eV Fermi level shift. Intrinsic defects or proper Hubbard corrections may reposition the Fermi level, potentially restoring semiconducting behavior experimentally.

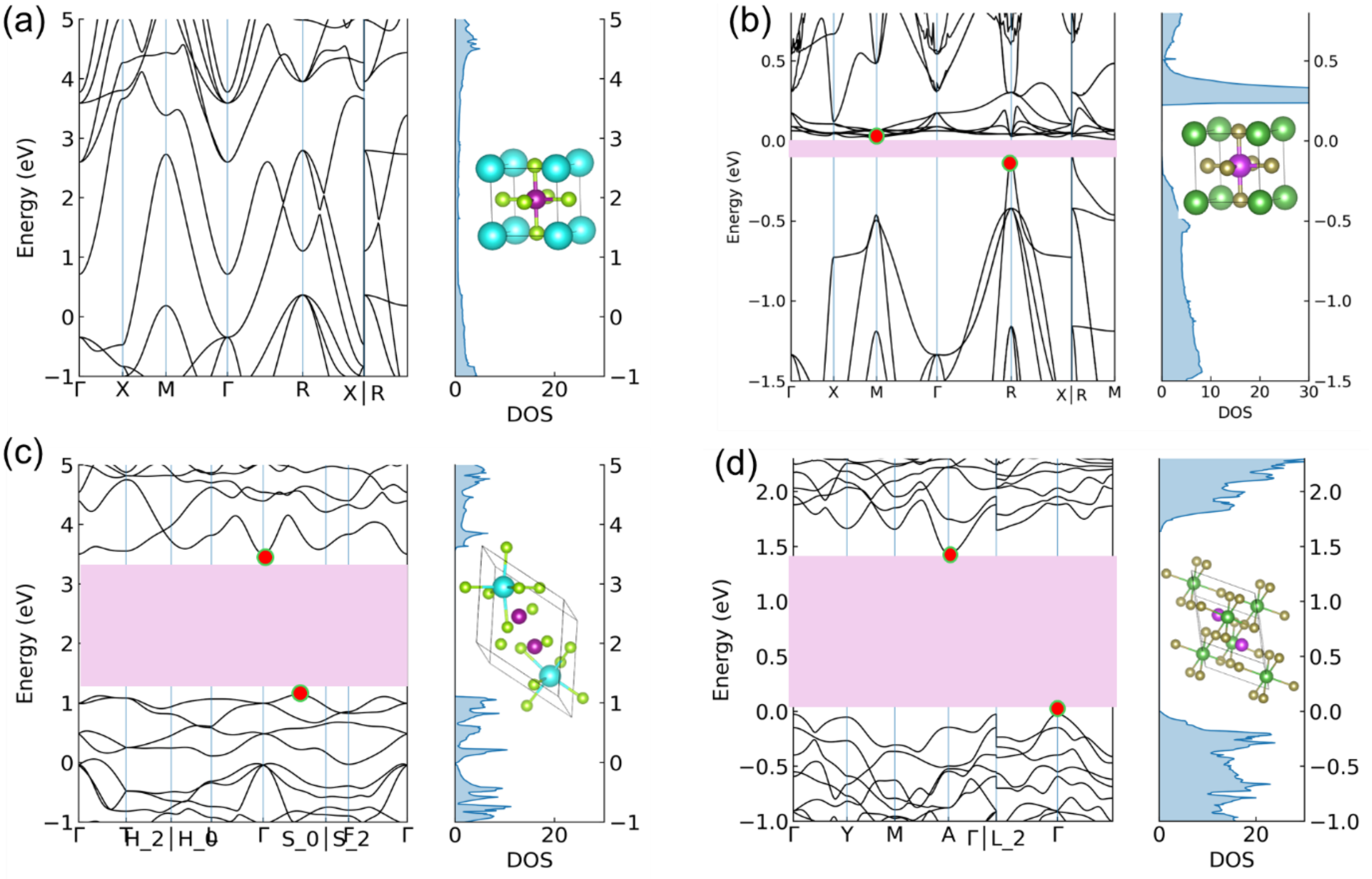

**Figure S5.** Band structure of the (a) $YbMnSe_3$ and (b) $LaBiTe_3$ and after the VibroML treatment (c) and (d), respectively.

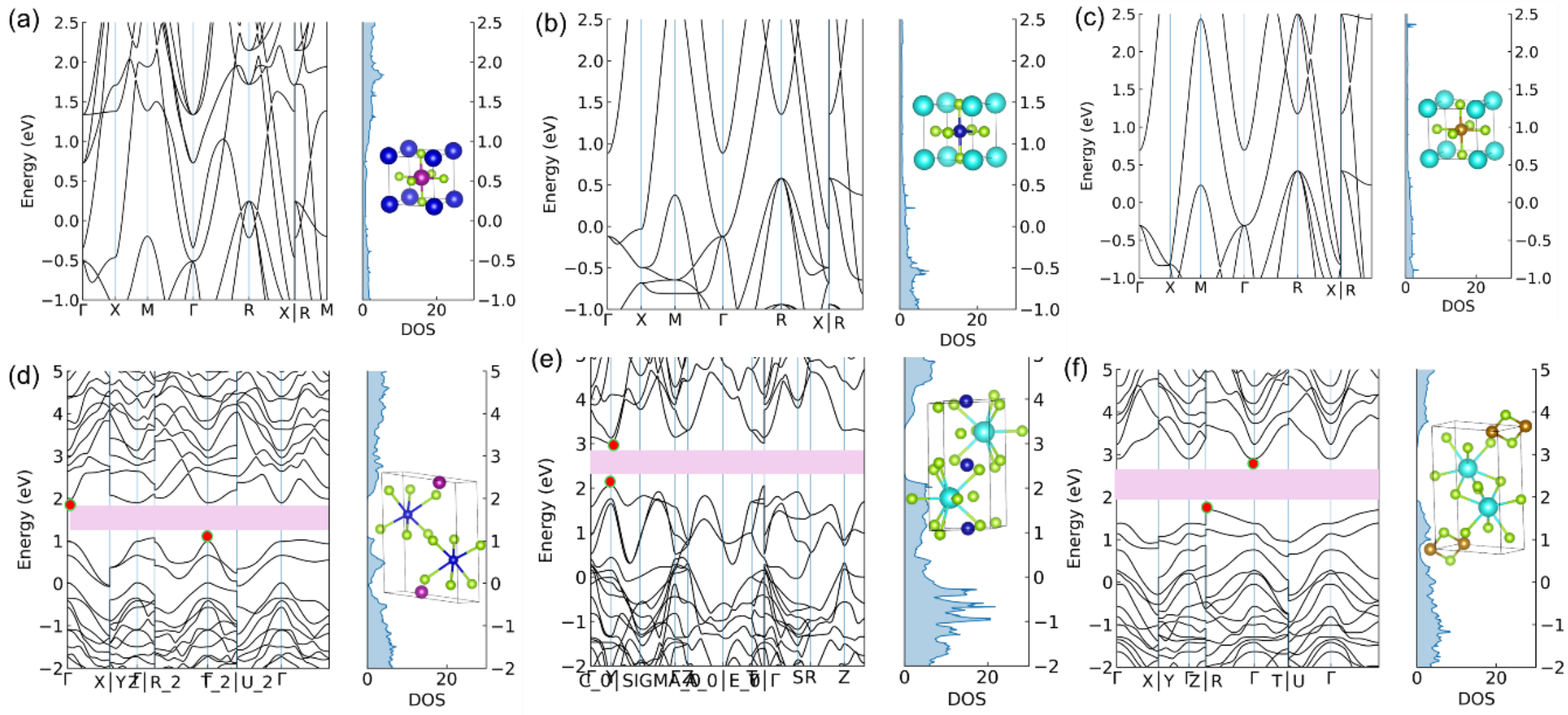

**Figure. S6.** Band structure of the (a) $TmMnSe_3$, (b) $YbCoSe_3$ and (c) $YbFeSe_3$ and after the VibroML treatment (d), (e) and (f), respectively.

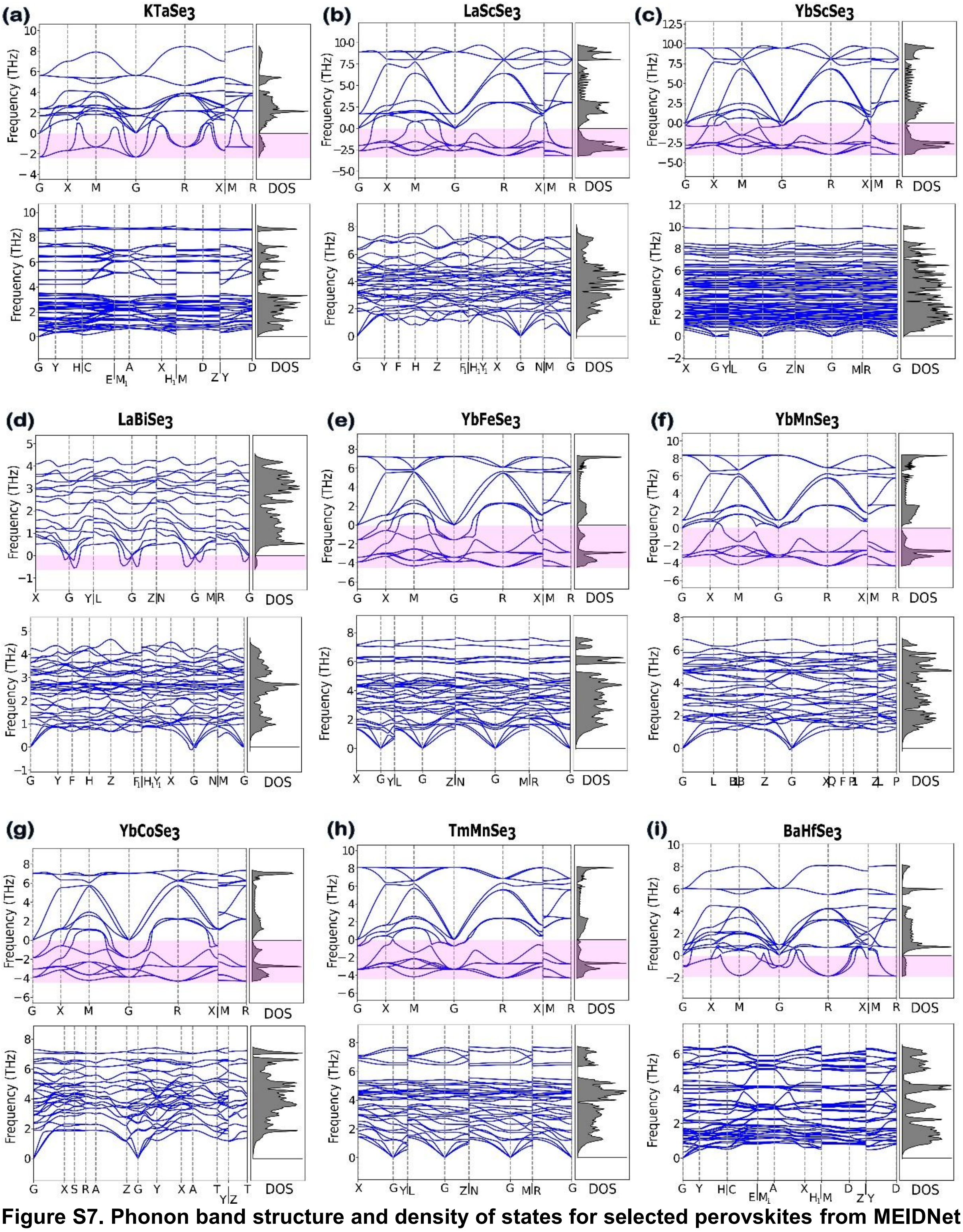

**Figure S7. Phonon band structure and density of states for selected perovskites from MEIDNet workflow obtained with MACE-OMAT-0 potential.** Each panel (a-i) compares the original structure with highlighted soft modes in pink (top) to the corresponding stabilized structure after remediation by VibroML (bottom).

**Associate Discussion S1: Early fusion and Late fusion**

We study two multimodal training schemes for crystalline materials: an EGNN-based crystal encoder–decoder (structure modality) coupled with scalar property modalities formation enthalpy and band gap. Both schemes project modality into a shared latent space and use a contrastive objective (InfoNCE) to align them, in the spirit of CLIP-style multimodal training.

**S1.1 Early fusion**: It concatenates the two scalars (bandgap and formation enthalpy) first and learns a single property latent. This reduces parameters and contrastive pairs, acting as a regularize in low-data or high-noise regimes [S1-3]. A curriculum factor $w_t$ (linear warm-up) stabilizes optimization by gradually increasing the contrastive weight up to 1 up to two-thirds of total epochs, and when there is no curriculum learning the weight is fixed to 1, during the alignment:

$$z_p = f_p([h; d]) \quad \text{(i)}$$

$$z_{joint} = (y_c + y_p)/2 \quad \text{(ii)}$$

$$L_{early} = L_{recon} + \left((\hat{h} - h)^2 + (\hat{d} - d)^2\right) + w_t * \lambda * NCE(c, p) \quad \text{(iii)}$$

**S1.2 Late fusion**: It uses *separate encoders* for *bandgap* and *formation enthalpy*, which computes pairwise contrastive alignment among, and averages the three normalized embeddings to form a joint latent for decoding [S4-6]. This triangulation typically yields a tighter shared space and robustness when one property is noisy or partially missing, and it scales to additional properties by adding encoders/decoders without redesigning the fused path. The contrastive alignment follows InfoNCE, widely used for representation learning across modalities:

$$z_{joint} = (y_c + y_h + y_d)/3 \quad \text{(iv)}$$

$$L_{prop} = (\hat{h} - h)^2 + (\hat{d} - d)^2 \quad \text{(v)}$$

$$L_{late} = L_{recon} + L_{prop} + \lambda * (1/3) * \left(NCE(c, h) + NCE(c, d) + NCE(h, d)\right) \quad \text{(vi)}$$

Where, h: formation enthalpy; d: band gap, $\hat{h}, \hat{d}$: predicted properties; $y_*$: projected and L2-normalized embeddings, $z$: latents, $z_{joint}$: joint latent used for decoding; $f_p$: property encoder; $L_{recon}$: crystal reconstruction loss; $L_{prop}$: property regression loss, NCE(·,·): contrastive InfoNCE; $\lambda$: contrastive weight; $w_t$: curriculum factor.

**Associate Discussion S2: Multimodal Generative Framework: From Bi- to Tri-Modality**

## S2.1 Data representation

Each crystal is encoded as a dense vector

$$v \in R^{6+N^2+N|S|+3N} \quad \text{(vii)}$$

6: lattice parameters $\boldsymbol{\ell} = (a, b, c, \alpha, \beta, \gamma)$

$N^2$: flattened adjacency $A \in \{0,1\}^{N\times N}$(neighbors within cutoff $r_c = 4$ Å)

$N|S|$: species one-hot matrix $S \in \{0,1\}^{N\times|S|}$over an element set $S$ with $|S| = 56$

$3N$: fractional coordinates $X \in [0,1]^{N\times 3}$

In the bimodal setting the second modality is a scalar property $y \in R$ (e.g., band gap). In the tri-modal setting we use two scalar properties $y = [y_1, y_2] \in R^2$ (e.g., $y_1 =$ band gap, $y_2 =$ formation enthalpy).

### S2.2 *E(3)-equivariant crystal encoder/decoder*

We use an E(3)-Equivariant GNN (EGNN) backbone [1]. For node (atom) $i$, let hidden features be $h_i \in R^{d_h}$ and coordinates $x_i \in R^3$. With adjacency $A_{ij} \in \{0,1\}$ and relative squared distance $d_{ij}^2 = |x_i - x_j|_2^2$, one layer performs:

$$e_{ij} = \phi_e([h_i, h_j, d_{ij}^2]) \quad \text{(viii)}$$

$$\Delta x_i = \sum_j A_{ij}(x_i - x_j)\phi_x(e_{ij}), \quad x_i' = x_i + \Delta x_i \quad \text{(ix)}$$

$$h_i' = \phi_h([h_i, \sum_j A_{ij} e_{ij}]) \quad \text{(x)}$$

where $\phi_e, \phi_x, \phi_h$are MLPs; $e_{ij}$ is an edge embedding; $d_h$ is hidden feature dimension. We center coordinates per structure $(x_i \leftarrow x_i - \bar{x})$ to ensure translation handling; updates depend only on relative geometry, yielding E(3)-equivariance. 3 EGNN blocks are used in encoder and decoder.

The crystal encoder $f_c: \mathbf{v} \mapsto \mathbf{z}_c \in \mathrm{R}^{128}$aggregates masked node features with a lattice MLP.

The crystal decoder $d_c: \mathrm{z} \mapsto (\hat{\ell}, \hat{A}, \hat{S}, \hat{X})$predicts lattice, adjacency logits, species logits, and coordinates. During decoding we assume a fully connected graph; final adjacency uses a logistic sigmoid

$$\hat{A} = \sigma\,(logits_A) \quad \text{(xi)}$$

$$\sigma(u) = \frac{1}{1+e^{-u}} \quad \text{(xii)}$$

and discrete species are obtained with a straight-through Gumbel-Softmax sampler with temperature $\tau > 0$

## S2.3 Property branches and common latent alignment

### S2.3.1 *Bi-modality (structure and band gap)*

A property encoder $f_p: R \to R^{128}$ maps $y$ to $z_p$. Projection heads $g_c, g_p: R^{128} \to R^{128}$ map to a common unit-sphere latent:

$$\widetilde{z_c} = \frac{g_c(z_c)}{|g_c(z_c)|_2}, \qquad \widetilde{z_p} = \frac{g_p(z_p)}{|g_p(z_p)|_2} \qquad \text{(xiii)}$$

$$z = (1/2)\left(\widetilde{z_c} + \widetilde{z_p}\right) \qquad \text{(xiv)}$$

The decoder $d_c$reconstructs the crystal from $\mathbf{z}$. For stability in bi-modality we also use the regression of scalar from the crystal branch via

$$\hat{y} \; = \; d_p{}^{(bi)}(z_c) \qquad \text{(xv)}$$

$$d_p^{(\mathrm{bi})}: R^{128} \to R \qquad \text{(xvi)}$$

### S2.3.2 *Tri-modality (structure + bandgap + formation enthalpy)*

We use the early fusion the two scalars $y = [y_1, y_2]$ through a property encoder $f_{pp}: R^2 \to R^{128}$ to obtain $z_{pp}$, then project as in Eq. (4):

$$\widetilde{\boldsymbol{z}_{pp}} = \frac{g_{pp}(\boldsymbol{z}_{\boldsymbol{pp}})}{|g_{pp}(\boldsymbol{z}_{\boldsymbol{pp}})|_2} \qquad \text{(xvii)}$$

$$z = (1/2)\left(\widetilde{z_c} + \widetilde{z_{pp}}\right) \qquad \text{(xviii)}$$

$z_{pp}$: fused latent of two properties [y1, y2] after an encoder.

$g_{pp}$: projection head for the fused-property branch.

In the tri-modal case, the property decoder $d_p^{(\mathrm{tri})}: R^{128} \to R^2$ predicts $[\widehat{y_1}, \widehat{y_2}]$ from the joint $z$.

## S2.4 Losses

### S2.4.1 *Reconstruction loss over lattice, adjacency, species, and coordinates*

$$L_{rec} = \left|\left|\hat{\mathrm{l}} - \mathrm{l}\right|\right|_2^2 + BCELogits(\hat{A}, A) + CE(\hat{S}, S) + \left|\left|\hat{X} - X\right|\right|_2^2 \qquad \text{(xix)}$$

$L_{rec}$: reconstruction loss for crystal modality.

$\hat{\ell}, \hat{A}, \hat{\mathrm{S}}, \hat{X}$: predicted lattice, adjacency, species logits (via Gumbel-Softmax), and coordinates.

$BCELogits$: binary cross-entropy with logits (for adjacency).

$CE$: cross-entropy (for species classification).

**S2.4.2 *Mean-squared error for bi-modality and tri-modality***

$$L_{prop}^{(bi)} = ||\hat{y} - y||_2^2, \quad \hat{y} = d_p^{(bi)}(z_c) \tag{xx}$$

${L_{prop}}^{(bi)}$: property loss in bi-modality (single scalar).

$d_p^{(bi)}$: regression head predicting y from $z_c$.

ŷ: predicted scalar property; y: ground truth.

$$L_{prop}^{(tri)} = \left|\left|\hat{\vec{y}} - \vec{y}\right|\right|_2^2, \quad \hat{\vec{y}} = d_p^{(tri)}(z) \tag{xxi}$$

${L_{prop}}^{(tri)}$: property loss in tri-modality (two scalars).

$-d_p^{(tri)}$: regression head predicting [y1, y2] from joint latent z.

$\vec{\hat{y}}$: predicted property vector; $\vec{y}$: ground-truth vector.

**S2.4.3 *Symmetric InfoNCE contrastive loss aligning structure and property embeddings via cosine similarity***

$$\mathcal{L}_{\text{InfoNCE}} = -\frac{1}{B}\sum_{k=1}^{B} \log \frac{\exp\left(\text{sim}\left(\boldsymbol{z_s^{(k)}}, \boldsymbol{z_p^{(k)}}\right)/\tau\right)}{\sum_{l=1}^{B} \exp\left(\text{sim}\left(\boldsymbol{z_s^{(k)}}, \boldsymbol{z_p^{(l)}}\right)/\tau\right)} \tag{xxii}$$

where $B$ is the batch size, and indices $(k)$ and $(l)$ denote samples within the mini-batch. $z_s^{(k)}$ and $z_p^{(k)}$ are the aligned structural and property embeddings for the $k$-th crystal, $sim(u, v) = u^{\mathrm{T}}v/(||u||||v||)$ denoting cosine similarity. τ is the temperature hyperparameter, and the denominator sums over all $l$ samples in the batch to normalize the probability.

**S2.5 Total loss**

**S2.5.1 Total objective for bi-modality: reconstruction + one property (bandgap) + curriculum-weighted contrastive term**

$$L = w_{rec}\, L_{rec} + w_{prop}\, L_{prop}^{(bi)} + w_{con}(t)\, L_{con} \tag{xxiii}$$

**S2.5.2 Total objective for tri-modality: reconstruction + two-property (bandgap and formation enthalpy) + curriculum-weighted contrastive term**

$$L = w_{rec}\, L_{rec} + w_{prop}\, L_{prop}^{(tri)} + w_{con}(t)\, L_{con} \tag{xxiv}$$

$w_{rec}, w_{prop}$ : fixed weights; $w_{con}(t)$: time-dependent contrastive weight